\documentclass[aps,twocolumn,amsmath,amssymb,prl,superscriptaddress,floatfix]{revtex4-2}

\usepackage{bm}
\usepackage{graphicx}
\usepackage{dcolumn}

\begin{document}

\title{Effects of scattering on the field-induced $T_c$ enhancement in thin superconducting films in a parallel magnetic field}

\author{V. G. Kogan}
  \affiliation{Ames National Laboratory - DOE, Ames, IA 50011, USA}
    \author{R. Prozorov}
    \email[Corresponding author: ]{Prozorov@ameslab.gov}
  \affiliation{Department of Physics \& Astronomy, Iowa State University, Ames, IA 50011, USA}
  \affiliation{Ames National Laboratory - DOE, Ames, IA 50011, USA}

  \begin{abstract}
The problem of a thin superconducting film in a parallel magnetic field, first discussed in the classical paper by Ginzburg and Landau for temperatures close to the critical, is revisited with the help of the microscopic BCS theory for arbitrary temperatures taking pair-breaking and transport scattering into account. While confirming experimental findings of the $T_c$ enhancement by the magnetic field, we find that the transport scattering pushes the phase transition curve to higher fields and higher temperatures for nearly all practical scattering rates. Still, the $T_c$ enhancement disappears in the dirty limit. We also consider intriguing changes, such as re-entrant superconductivity, caused to the phase boundary by pair-breaking magnetic ions spread on one of the film faces. These features await experimental verification.
 \end{abstract}
\date{\today }

\maketitle

\section{Introduction}

Thin films are major elements of superconducting devices, such as bolometers, SQUIDs, fault-current limiters, and qubits for quantum computers. Their physical properties in the superconducting state determine the device performance metrics, such as quantum coherence. Unexpected properties of films in the parallel magnetic field were recorded \cite{Goldman}.
The enhancement of superconductivity (SC) in thin films in parallel fields is now an established experimental fact \cite{Xiong, WTe2}. A few nontrivial mechanisms for this enhancement were suggested \cite{WTe2,Feig}. However, the opinion expressed in \cite{Xiong} leans toward early work \cite{K86,KN87}, which describes this phenomenon as a consequence of the ``bare" classical weak-coupling BCS theory.

The main point of this interpretation is that  at the 2nd order phase transition, the order parameter satisfies a linear equation (as in Helfand and Werthamer  treatment of the upper critical field $H_{c2}(T)$ \cite{HW})

\begin{eqnarray}
{\bm \Pi}^2\Delta =k^2\Delta\,,
\label{HW}
\end{eqnarray}
where ${\bm  \Pi} =\nabla +2\pi i{\bm
A}/\phi_0$ with    the vector potential $\bm A$ and  the flux quantum $\phi_0$.
In fact, this equation holds at any 2nd order transition from  normal to  SC state away of $H_{c2}$, e.g. in proximity systems or at $H_{c3}$,  provided $k^2=-1/\xi^2$ satisfies the self-consistency equation of the theory \cite{K85,KN87}. It turned out that the coherence length $\xi$ so evaluated depends not only on temperature and scattering but also on the magnetic field (except in the dirty limit or near $T_c$). The field dependence has been confirmed in Scanning Tunneling measurements of the length scale of spatial variation of $\Delta$ \cite{Jesus}, in $\mu$SR data  \cite{muSR}, and in data on macroscopic magnetization $M(H)$ \cite{M(H)} for a number of materials.

Solving Eq.\,(\ref{HW}), one   imposes certain boundary conditions on the order parameter $\Delta$. In the bulk problem of $H_{c2}$, $\Delta(\bm r)$ should be finite everywhere, in the problem of $H_{c3}$ for nucleation of SC at the sample surface one requires $\partial_x\Delta=0$ at the sample surface ($x$ is normal to the surface \cite{DG}), for a thin film in the parallel field this gradient is required to vanish on both film faces \cite{KN87}.

In this work, we extend  the earlier treatment \cite{KN87} by including pair-breaking scattering having in mind possible interpretations for experimental results described in \cite{Xiong}. Besides, we consider the effect of replacement of the condition $\partial_x\Delta=0$ at one of the film surfaces with $ \Delta=0$ to describe the pair-breaking by magnetic ions spread at this surface (in a manner described in \cite{Xiong}). The resulting phase transition curves have several surprising and unexpected features, which demonstrate an extreme sensitivity of these curves to boundary conditions.

    Consider an isotropic material with both magnetic and non-magnetic scatterers; $\tau_m$ and $\tau$ are the corresponding average scattering times. The problem of the second order phase transition from the normal to SC phases can be addressed based on Eilenberger quasiclassical
version of Gor'kov's equations for normal and anomalous Green's
functions $g$ and $f$. At the 2nd order phase transition, $g=1$ and we are left with a linear equation for $f$ \cite{Eil,KPMishra}:
\begin{eqnarray}
(2\omega^+ +{\bf v}\cdot {\bm \Pi})\,f=2\Delta
/\hbar+ \langle f\rangle/\tau^- \,,\label{E1}\\
\omega^+=\omega + \frac{1}{2\tau^+}\,,\qquad\frac{1}{\tau^\pm}=\frac{1}{\tau
}\pm\frac{1}{\tau_m}\,.
\label{om+}
\end{eqnarray}
Here, ${\bm v}$ is the Fermi velocity,  $\Delta ({\bm  r})$ is  the order parameter;
  Matsubara frequencies are defined by $ \omega=\pi T(2n+1)$
with an integer $n$; in the following (except  some final results) we set $\hbar=k_B=1$; $\langle...\rangle$ stand for  averages over the Fermi surface.
Solutions $f$ of  Eq.\,(\ref{E1}) along with   $\Delta$ should satisfy the  self-consistency equation:
\begin{equation}
\frac{\Delta}{2\pi  T}\ln\frac{T_{c0}}{T}=\sum_{\omega>0}\left(\frac{\Delta}
{ \omega }-\langle f\rangle\right)\,,
\label{selfcons0}
\end{equation}
where $T_{c0}$ is the critical temperature in the absence of pair-breaking scattering.

Repeating the derivation  of Ref.\,\cite{K85}, one finds (see the outline in Appendix A):
 \begin{equation}
\langle f\rangle = \Delta\, \frac{2\tau^- S}{2\omega^+\tau^- -S} \,,
\label{F}
\end{equation}
where $S$ is given by a series
\begin{eqnarray}
S &=&\sum_{j,m=0}^\infty \frac{(-q^2)^j}{j!(2m+2j+1)}
\left(\frac{(m+j)! }{m!}\right)^2
\left(\frac{\ell^+ }{\beta^+}\right)^{2m+2j}
\nonumber\\
&\times&  \prod_{i=1}^{m}\left[k^2+(2i-1)q^2\right],\qquad q^2=\frac{2\pi H}{\phi_0}\,,
 \label{Series}
\end{eqnarray}
where
\begin{eqnarray}
  \ell^+=v\tau^+,\qquad  \beta^+=1+2\omega \tau^+\,.
\label{ell+}
\end{eqnarray}

This sum can be transformed to an integral, which is more amenable for the numerical work \cite{KN87}:
\begin{eqnarray}
S&=& \sqrt{\frac{\pi}{u}} \int_0^1\frac{d\mu\,(1+\mu^2)^\sigma}{(1-\mu^2)^{\sigma+1}}
 \left[{\rm erfc}\frac{\mu}{\sqrt{u}}-\cos(\pi\sigma) {\rm erfc}\frac{1}{\mu\sqrt{u}}\right], \nonumber\\
u&=& \left( \frac{q \ell^+}{\beta^+}\right)^2.
\label{SKN}
\end{eqnarray}
  Introducing dimensionless quantities:
 \begin{equation}
h=H  \frac{2\pi d^2}{\phi_0}=q^2d^2,\,\,\,\, P^\pm=\frac{\hbar}{2\pi T_{c0}\tau^\pm} =P \pm P_m \qquad
\label{variables}
\end{equation}
 and the reduced thickness
\begin{eqnarray}
D =d\,\frac{2\pi T_{c0}}{\hbar v}  \,,
 \label{1/r}
\end{eqnarray}
one obtains
\begin{eqnarray}
u= \frac{h}{D^2[P^++t(2n+1)]^2}\,.
  \label{u}
\end{eqnarray}

The parameter $\sigma$ as defined in \cite{KN87} is
\begin{eqnarray}
    \sigma=\frac{1}{2}\left(\frac{k^2}{q^2}-1\right) \,.
    \label{sig}
\end{eqnarray}
 This parameter depends on the   phase transition in question: it is easy to see that $\sigma=-1$ at $H_{c2}(T)$. For $H_{c3}$ near $T_c$ of a half-space sample the result of Saint-James and DeGennes leads to $\sigma=-0.795$ \cite{DG}; transport scattering leads to the temperature dependence of $\sigma$ \cite{Levch,crit-fields}.


For  numerical work we recast the self-consistency relation (\ref{selfcons0}) combined with Eq.\,(\ref{F}) to dimensionless form:
\begin{eqnarray}
-\ln t=  \sum_{n=0}^\infty\left[ \frac{1}{n+1/2}-\frac{2tS}{2t(n+1/2)+P^+-SP^-}\right]\qquad\nonumber\\
\label{s-c}
\end{eqnarray}
with the reduced temperature $t=T/T_{c0}$.

\section{Symmetric boundary conditions $\bm {\Delta^\prime(\pm d/2) =0}$ }

  As mentioned above, the order parameter at a 2nd order  phase transition satisfies $\Pi^2\Delta=k^2\Delta$.  Choose the plane $(y,z)$ parallel to the film and  $x=0$ in the film middle.  Denoting
\begin{equation}
   s=qx,\qquad \eta=-k^2/q^2 \,,
\label{eqDel}
\end{equation}
 we obtain a differential equation
\begin{equation}
\Delta''(s)-s^2\Delta(s)=-\eta\Delta (s)\,,
\label{eqDel}
\end{equation}
so $-\eta$ is the eigenvalue of the linear operator at the LHS.
The general solution   is:
 \begin{eqnarray}
\Delta &=&e^{-s^2/2}\left[_1F_1\left(\frac{1-\eta}{4},\frac{1}{2},s^2\right)+Cs\, _1F_1\left(\frac{3-\eta}{4},\frac{3}{2},s^2\right)\right]\nonumber\\
 \label{solution}
\end{eqnarray}
 where $_1F_1$  are   confluent hypergeometric functions and $C$ is an arbitrary constant.
The symmetry with respect to the film middle gives $C=0$, and   the condition $\Delta'(\pm d/2)=0$ yields
 \begin{equation}
(1-\eta)\, _1F_1 \left(\frac{5-\eta}{4}, \frac{3}{2}, \frac{ h }{4}\right) =\,_1F_1 \left(\frac{1-\eta}{4}, \frac{1}{2}, \frac{h}{4}\right) \,.
\end{equation}
 Hence, for a given field $h$, the eigenvalue $\eta$ can take only a certain value, the root of this equation.

  Given $\eta(h)$, we evaluate $\sigma=(k^2/q^2-1)/2=-(\eta+1)/2$ for this value of $h$, and, therefore, we can calculate $S(u,\sigma)=S(h,t,n)$ for given $P$, $P_m$, $D$ and solve the self-consistency equation for $t$ at this $h$. Scanning $h$ we  recover the whole transition curve curve  $t(h)$. In general, scanning $t$ would not work because $h(t)$ might happen to be multi-valued.

\begin{figure}[tbh]
\includegraphics[width=7cm]{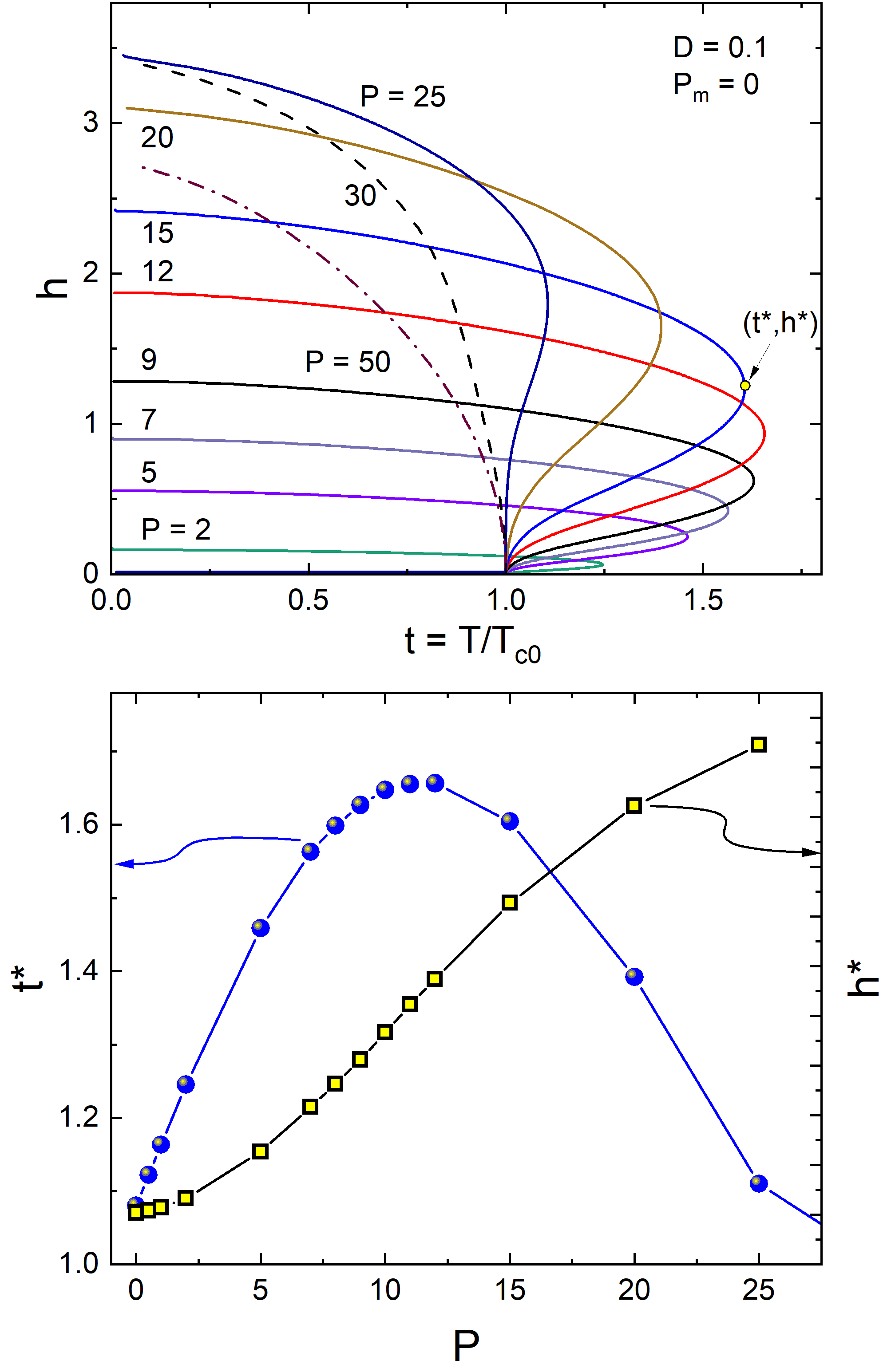}
 \caption{(Color online) The upper panel: the phase boundary $h(t)$  for a set of different transport scattering rates $P$ indicated. The position of the maximum enhancement $(t^*,h^*)$ is shown for one curve, $P=$15. The lower panel: the position of the maximum enhancement  $t^*$ (left axis) and $h^*$ (right axis) as function of $P$. On both panels $D=0.1$, $P_m=0$. }
 \label{fig1}
 \end{figure}

 The numerical results for purely transport scattering are shown in Fig.\,\ref{fig1}.
It is worth noting that the transport scattering causes an increase of the $T_c$-enhancement up to $P\sim 10$ and only for strong scattering with $P> 10$ it suppresses the effect in agreement with the general theoretical statement that the effect should disappear in the dirty limit \cite{K85,KN87}.

Basically, the physical reason for the enhancement of the superconducting region on the $(T,H)$ phase diagram by transport scattering is the same as one for the case of enhancement of the bulk upper critical field $H_{c2}(T)$: the scattering caused reduction of the coherence length $\xi(T,P)$ yet enforced by the field dependence of $\xi$.

\begin{figure}[tbh]
\includegraphics[width=7cm]{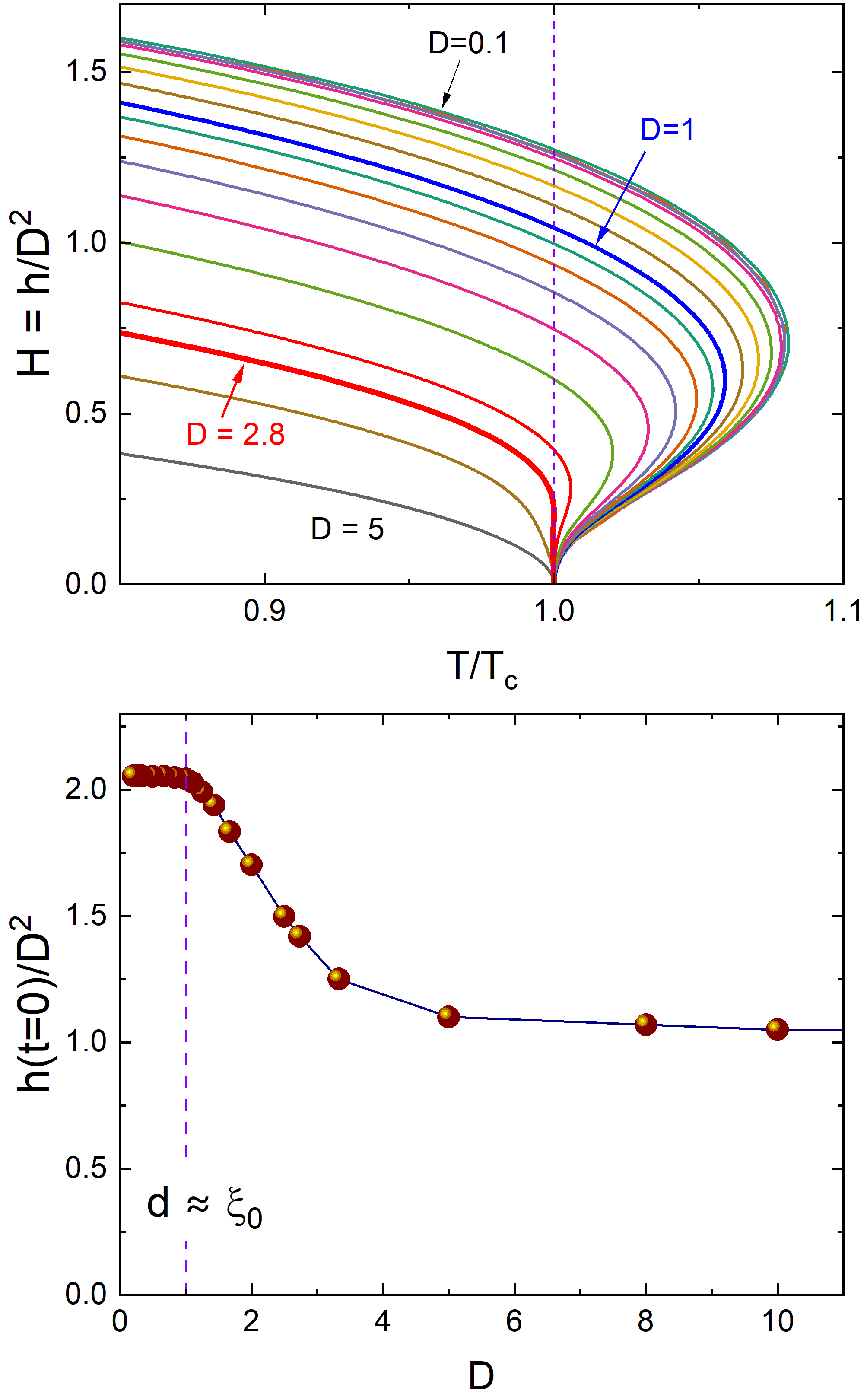}
 \caption{(Color online) The upper panel: the clean-limit phase boundary $h(t)$ zoomed at high temperatures for a set of thicknesses indicated.
The $T_c$ enhancement disappears at approximately $D=$2.8 The lower panel: the close-to-zero-temperature  field $ h(0)/ D^2\propto H(0)$ vs. $D$. $P=P_m=0$. }
 \label{fig2}
 \end{figure}

The upper panel of Fig.\,\ref{fig2} shows  phase boundaries for the clean case and a set of thicknesses from $D=0.1$ to $D=5$ at  temperatures close to $T_{c0}$. The maximum $D$ with a finite $T_c$-enhancement in this set is $D=2.8$, close to $\approx 2.6$,   predicted based on GL theory \cite{KN87}.

The lower panel shows that at $T $ close to zero, the field
 \begin{equation}
H=\frac{h}{D^2}\, \frac{2\pi\phi_0 T_{c0}^2}{\hbar^2v^2}\,
\label{H}
\end{equation}
(in common units)   is nearly constant for small thicknesses in the interval $0.1<D\lesssim 1$, but it decreases for thicker films. One should have in mind that the ``laminar" structure of $\Delta(x)$ of thin films with increasing thickness becomes unstable and gives way to vortices with $\Delta(x,y)$ as had been shown in \cite{Fink,Schult} within GL theory. Although the large values of $D$ are irrelevant for the film problem, it is interesting to note that $H(0)$ for large $D$ is close to $H_{c2}(0)$.

\begin{figure}[tbh]
\includegraphics[width=7cm]{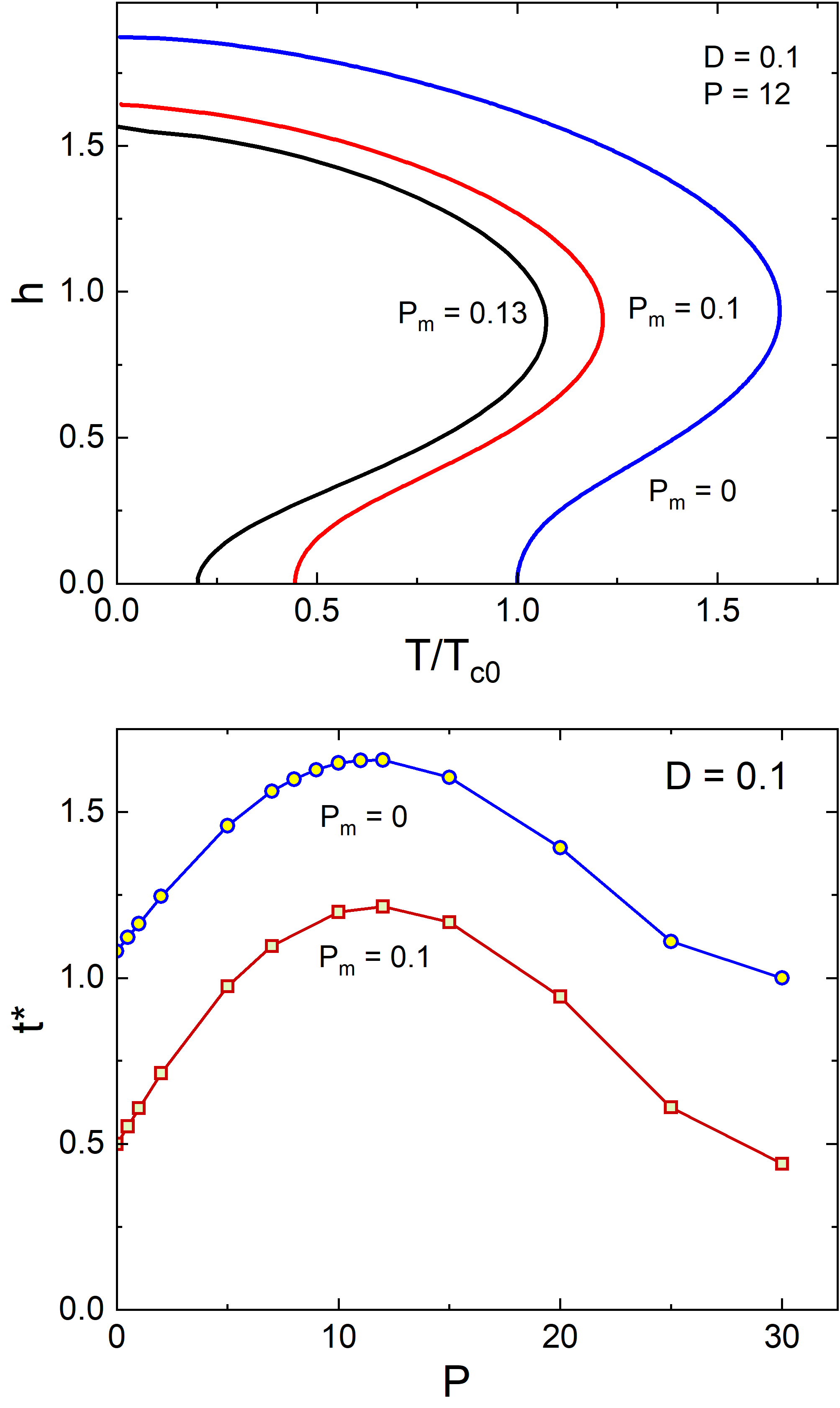}
 \caption{(Color online) The upper panel: the phase boundary $h(t)$ for different magnetic scattering rates,  $P_m$, at a fixed $P=12$. The power panel: the maximum $t_c$ enhancement, $t^*$ vs. transport scattering rate $P$ for two fixed $P_m$. At both panels $D=0.1$. }
 \label{fig3}
 \end{figure}

Figure \ref{fig3} shows that the magnetic scattering, while strongly suppressing the critical temperature, leaves    the $T_c$ enhancement effect, i.e. $(t^*-t_c)$ nearly unchanged. For curves $P_m=0$ and $P_m=0.13$, $T_c$ drops by a factor of 5, whereas $(t^*-t_c)$ changes by about $30\%$.

 It should be noted that the scattering parameters $P$ and $P_m$ refer to the bulk properties of the film material. Effects of magnetic ions on one of the film faces can be taken into account by the boundary conditions at the surfaces rather than by the value of $P_m$, the subject of the next Section.

\section{Mixed boundary conditions: $\bm{\Delta^\prime( d/2)=0}$  on one surface and $\bm {\Delta(- d/2)=0}$ on the other}

These conditions can be realized in a film on an insulating substrate ($\Delta^\prime=0$) with magnetic ions spread at the other surface ($\Delta=0$).
We have chosen these boundary conditions to demonstrate how sensitive the phase boundary in films placed in the parallel field is to the film environment. In fact, Saint-James and de Gennes   pointed this out  in their seminal work \cite{DG}.

 The boundary conditions of the Section title suffice to determine both the arbitrary constant $C$ and the parameter $\eta$ of the general solution (\ref{solution}).
 $\Delta(- d/2)=0$ yields
 \begin{eqnarray}
C  =\frac{2}{\sqrt{h}}\,_1F_1\left(\frac{1-\eta}{4},\frac{1}{2},\frac{h}{4}\right)\Big/ \,_1F_1\left(\frac{3-\eta}{4},\frac{3}{2},\frac{h}{4}\right).\qquad
  \label{C}
\end{eqnarray}
The condition $\Delta^\prime( d/2)=0$  results in:
 \begin{eqnarray}
&& 3 C\left(1+\frac{h}{4}\right)\, _1F_1\left(\frac{3-\eta}{4},\frac{3}{2},\frac{h}{4}\right)  \nonumber\\
&& + \frac{\sqrt{h}}{2}\Big[3\,_1F_1\left(\frac{1-\eta}{4},\frac{1}{2},\frac{h}{4}\right)
 -3(1+\eta)\,_1F_1\left(\frac{1-\eta}{4},\frac{3}{2},\frac{h}{4}\right) \nonumber\\
&&- C\frac{\sqrt{h}}{2}(3+\eta)\,_1F_1\left(\frac{3-\eta}{4},\frac{5}{2},\frac{h}{4}\right) \Big] =0.
  \label{2nd bc}
\end{eqnarray}
At a given $h$, Eqs.\,(\ref{C}) and (\ref{2nd bc})  can be solved for $\eta$ numerically ($\eta$  is needed   for the power $\sigma=-(\eta+1)/2$ in the integral $S$).

\subsection{Zero-field $\bm { T_c(d)} $ }

The  condition $\Delta(- d/2)=0$  suppresses the film $T_c$ even in the field absence. Evaluation of  this suppression  is necessary to interpret various transition curves $h(t)$.
 In zero field, the order parameter at the phase boundary satisfies   $\Delta^{\prime\prime}=k^2\Delta$ with the solution  $\Delta=\Delta_0\sin |k|(x+d/2)$ with $|k|=\pi/2d$.
    Further, $T_c$ should be found from the self-consistency Eq.\,(\ref{selfcons0}) which contains the quantity $S$ via Eq.\,(\ref{F}). The shortest way to get $S$ for $H=0$ is to go to the definition of $S$ as a power series \cite{K85,crit-fields} and set in it $q^2=2\pi H/\phi_0=0$:
   \begin{eqnarray}
 S=  \frac{1}{\mu} \,\tan^{-1}\mu\,,\qquad \mu=\frac{\pi}{2D}\,\frac{\ell^+}{\beta^+}\,.
  \label{So}
\end{eqnarray}

The self-consistency equation now reads:
   \begin{eqnarray}
-\ln t_c=  \sum_{n=0}^\infty\left[ \frac{1}{n+1/2}-\frac{2t_cS}{ t_c(2n+1 )+P^+-SP^-}\right]. \qquad
\label{s-c}
\end{eqnarray}
 where $t_c=T_c/T_{c0}$ and $ T_{c0}$ is the critical temperature of the bulk material in the absence of pair-breaking scattering.
The dimensionless parameter $\mu$ of Eq.\,(\ref{So}) is
   \begin{eqnarray}
  \mu=\frac{\pi}{2D}\,\frac{1}{t_c(2n+1 )+P^+ }\,,\qquad D =d\,\frac{2\pi T_{c0}}{\hbar v}\,,
  \label{mu,D}
\end{eqnarray}
so that one can numerically solve the self-consistency equation for $t_c(D)$.

 \begin{figure}[b]
 \includegraphics[width=7cm]{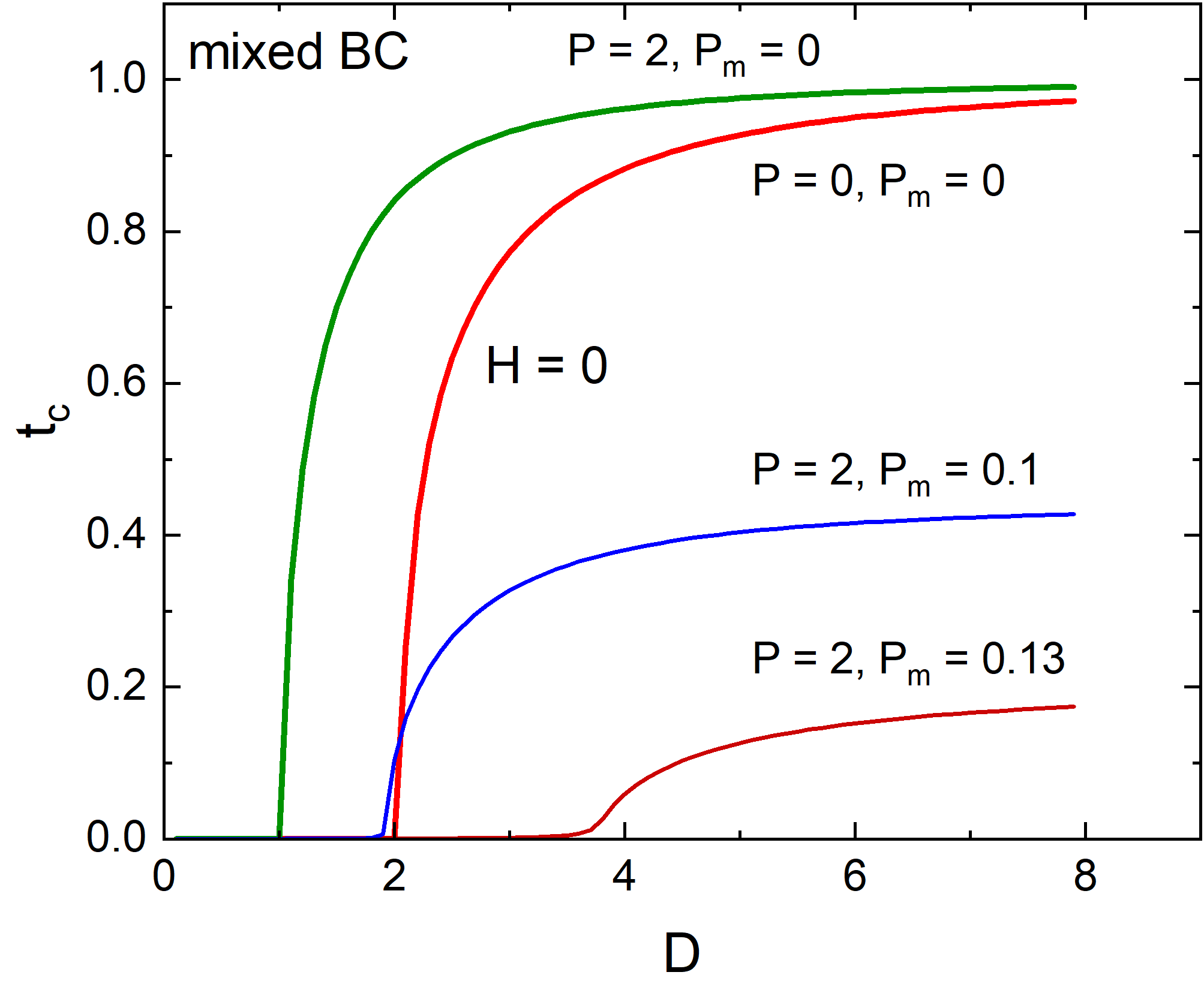}
 \caption{(Color online) Zero-field critical temperature $t_c(D)$ as a function of thickness $ D$ for the mixed boundary conditions and various combinations of the scattering parameters indicated.   }
 \label{fig4}
 \end{figure}

As the upper panel of Fig.\,\ref{fig4} shows, the requirement $\Delta=0$ at one of the film surfaces leads to a progressive reduction of $T_c$ with decreasing thickness. Moreover, $T_c$ turns zero at $D=D_c=2$ in the clean case, so that in zero field SC is completely suppressed for $D<D_c$. With increasing transport scattering, the sharp break of $t_c(D)$ at $D_c$ moves to smaller thicknesses.   
Hence, the phenomenology of SC films in parallel fields is quite peculiar.      The lower panel shows that the pair-breaking scattering smears the sharp break in $t_c(D)$ to a smooth crossover, whose position is shifted   to thicker films. Our example of $P_m=0.13$ corresponds to a strong pair breaking and the gapless situation in the bulk (recall that the critical value of $P_m$ where the bulk $T_c(P_m)=0$ is  $P_m=0.14$, see e.g. \cite{KPMishra}).

    It is instructive to observe that if $D\gg 1$, the parameter $\mu$ is small, and Eq.\,(\ref{s-c}) is reduced to the Abrikosov-Gor'kov bulk relation for $t_c(P_m)$:
     \begin{eqnarray}
-\ln t_c=  \psi\left(\frac{1}{ 2}+\frac{P_m}{t_c}\right)-\psi\left(\frac{1}{ 2}\right).
\label{AG}
\end{eqnarray}

\subsection{Phase boundary}

\begin{figure}[b]
\includegraphics[width=7.5cm]{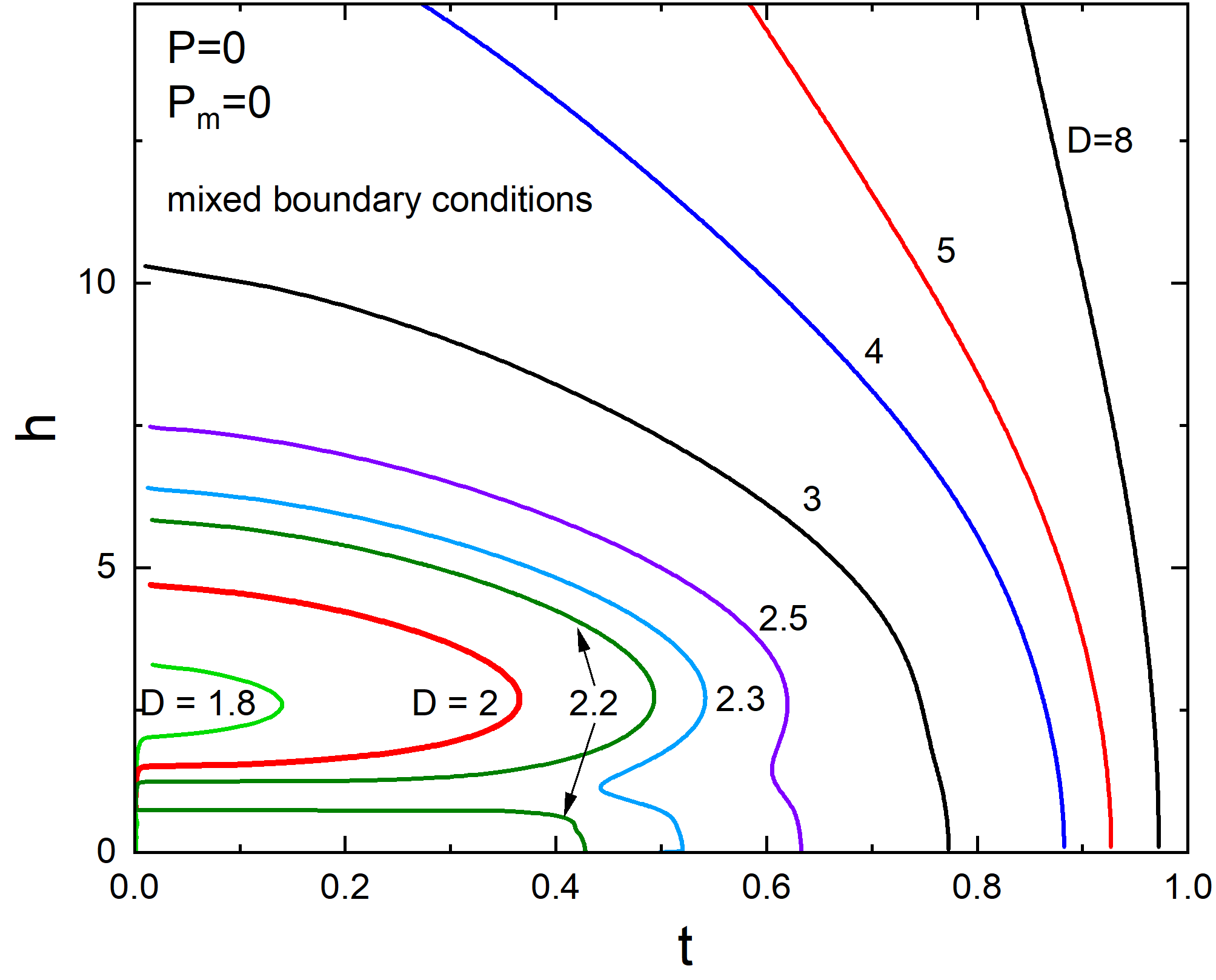}
 \caption{(Color online) The phase boundary $h(t)$ for the mixed boundary conditions on the order parameter, $ \Delta^\prime(D/2)=0$  on one surface, and $\Delta(- D/2)=0 $ on the other in a clean limit, for $P=0$, $P_m=0$. The reduced field is $h=(2\pi d^2/\phi_0)H $ and the reduced temperature $t=T/T_{c0}$. The numbers by the curves are dimensionless thicknesses $D=(2\pi T_{c0}/\hbar v) d$.
  }
 \label{fig5}
 \end{figure}

For thicknesses substantially exceeding $D_c$ ($D_c=2$ for the clean limit) the transition curve is of the type $H\sim\sqrt{T_c-T}$ with zero enhancement as shown in Fig.\,\ref{fig5}.

When  $D<D_c=2$, $T_c=0$ for $H=0$. Therefore, the low end of the transition curve should start at $t=0$, as is seen in Fig.\,\ref{fig5}   for  clean films with $D=1.8$ and 2.  The equilibrium SC does not exist at all out of these curves. For SC to exist   one should apply field within the area inside these  curves. In particular such a phase boundary implies that the magnetoresistance at a fixed temperature $ t\lesssim 0.36$ should have a minimum for $D=2$.  These examples demonstrate that thin films in parallel fields complement the list of phenomena where the magnetic field ``helps" SC instead of suppressing it, see e.g. \cite{Feig}.

Unlike the situation with $D<D_c=2$, for $D>D_c$ the zero-field $T_c$ is not zero, therefore the low end of the transition curve starts at $T_c>0$,  examples are shown in Fig.\,\ref{fig5} for   $D=2.2 $ and larger.
In the region of thicknesses adjacent to $D_c$,  the phase boundary may take a non-trivial shape shown for $D=2.2,\,\, 2.3,\,\, 2.5$  where
 the transition curve $h(t)$ becomes multi-valued.

\section{Summary}

 We have considered   thin superconducting films in a parallel magnetic field.
 Near $T_c$, this problem has been  discussed  in the classic GL paper \cite{GL}. Our approach is based on the quasi-classical microscopic theory of Eilenberger \cite{Eil} that holds for any $T$.

    We have shown that in the presence of impurities, the magnetic ones included, the model  \cite{K85,KN87} developed in the late 80's  still works. We obtained conditions for the remarkable effect recently observed \cite{Xiong}, the enhancement of the in-field critical temperature $T_c(H)$  above the zero-field $T_c(0)$.

 It turns out  that transport scattering amplifies this effect if the scattering parameter $P\sim \xi_0/\ell \lesssim 10$ ($\ell$ is the mean-free path and $\xi_0$ is the BCS zero-$T$ coherence length). For larger $P$, the $T_c$-enhancement is suppressed to disappear in the dirty limit \cite{K85,KN87}.
 This new insight clarifies the role of the transport scattering bringing it in line with the general theory prediction of absent $T_c$-enhancement in the dirty limit.
  On the other hand, this improves the chances to   observe the effect \cite{Xiong,WTe2}, because the transport scattering in thin films is usually strong. For example, the experiment \cite{Xiong} registered  $T_c$ enhancements in amorphous Pb films with the estimated mean-free path $\ell \approx 1\,$nm whereas estimates of $\xi_0$ in Pb range between $\approx 230$ and $300\,$nm.
In general, while in many modern superconductors, such as high-$T_c$ cuprates or is iron pnictides, $\xi_0$  is so short that makes it challenging to make such films, in technologically - important Nb it is quite feasible with $\xi_0 \approx 40$ nm.

  It should be noted that the effect of magnetic impurities {\it per-se} on the $T_c$-enhancement turned out relatively weak, the standard $T_c$ suppression notwithstanding, see Fig.\,\ref{fig3}.

The properties of thin films in a parallel magnetic field are very sensitive to physical conditions on their faces. If, for example, pair-breaking magnetic ions are spread over one of the faces while the opposite face   is on   an insulating substrate, then realistic boundary conditions for the order parameter would be $\Delta=0$ on one face and $\partial_x\Delta(x)=0$ at the other. This arrangement was in fact tested in experiment  \cite{Xiong}.

The major consequence of the condition  $\Delta=0$ on one of the faces is that it makes the critical temperature thickness dependent. We find that for $P=P_m=0$, the surface pair breaking kills the SC in zero-field for thicknesses $D<2$ (in units $\hbar v/2\pi T_{c0}$). But as Fig.\,\ref{fig5} shows, even for $D<2$, application of the magnetic field may cause the SC {\it reentrance} at a finite field interval. Hence, the film magnetoresistance at $T=\,$const should have a dip in this interval.

For $2<D\lesssim 3$, the competition of reentrance and finite $T_c$ causes the phase transition curve to acquire a non-trivial shape so that $h(t)$ becomes multi-valued, see Fig.\,\ref{fig5}.

It is worth mentioning that the problem of the phase boundary for a film in a parallel field is directly related to that of the surface critical field $H_{c3}$. In the latter case, SC near the surface can exist in fields out of the phase boundary of the bulk material. In the film case we have two surfaces close to each other and again SC can exist out of the bulk phase boundary in thin enough films. In particular, in finite fields it can exist above the zero-field critical temperature $T_{c0}$. The latter possibility, strange at first sight, is as ``strange"  as the SC in fields $H_{c2}<H<H_{c3}$, in both cases it exists out of the phase boundary of the bulk material.

\section{Acknowledgments}

This work was supported by the U.S. Department of Energy (DOE), Office of Science, Basic Energy Sciences, Materials Science and Engineering Division. Ames Laboratory is operated for the U.S. DOE by Iowa State University under contract \# DE-AC02-07CH11358.

RP acknowledges support from the DOE National Quantum Information Science Research Centers, Superconducting Quantum Materials and Systems Center (SQMS) under contract \# DE-AC02-07CH11359.

\appendix

\section{The sum $\bm S$ in the presence of magnetic impurities}

The solution $f$ of Eq.\,(\ref{E1}) can be written as
\begin{eqnarray}
 f &=&  (2\omega^+ +\bm v\bm\Pi)^{-1} (F/\tau^- + 2\Delta) \nonumber\\
 &=&   \int_0^\infty d\rho e^{-\rho(2\omega^+ +\bm v\bm\Pi)} (F/\tau^- + 2\Delta) \,.
\label{f}
\end{eqnarray}
Taking the Fermi surface average we get
\begin{eqnarray}
F=\frac{1}{\tau^-}   \int_0^\infty d\rho e^{-2 \omega^+\rho}\left\langle e^{-\rho\bm v\bm\Pi}\right\rangle(F+ 2\Delta\tau^- ) \,. \qquad
\label{Fa}
\end{eqnarray}
The term $\langle...\rangle$ does not contain the scattering parameters, hence it is the same as that calculated in \cite{K85,K86,KN87} for the clean case:
\begin{eqnarray}
\left\langle e^{-\rho\bm v\bm\Pi}\tilde F\right\rangle = \sum_{m,j}\frac{(-q^2)^j}{(m!)^2j!}\frac{(2\mu)!!}{(2\mu+1)!!} \left(\frac{\rho v}{2}\right)^{2\mu}\Pi^{+^m}\Pi^{-^m}\tilde F.  \nonumber\\
\label{average}
\end{eqnarray}
Here $\tilde F=F+2\Delta\tau^-$, $\mu=m+j$, and $\Pi^{\pm}=\Pi_x\pm i \Pi_y$.
After integrating over $\rho$, one obtains from Eq.\,(\ref{Fa})
\begin{eqnarray}
&&F=\frac{1}{2\omega^+\tau^-}
  \sum_{m,j}\frac{(-q^2)^j}{j!(2\mu+1)}\left(\frac{ \mu !}{m!}\right)^2 \left(\frac{\ell^+}{\beta^+}\right)^{2\mu} \Pi^{+^m}\Pi^{-^m}\tilde F \nonumber\\
&&\ell^+=v\tau^+,\quad  \beta^+=1+2\omega \tau^+\,.
\label{average1}
\end{eqnarray}
 One can check that if no magnetic impurities are involved, this reduces to Eq.\,(12) of \cite{K85}. Using commutation properties of operators $\Pi^\pm$ in uniform field, one manipulates
\begin{eqnarray}
 \Pi^{+^m}\Pi^{-^m}\tilde F = \tilde F \prod_{i=1}^{m} [k^2+(2i-1)q^2]
\label{algebra}
\end{eqnarray}
and obtains:
\begin{eqnarray}
F=\Delta \frac {2\tau^- S}{2\omega^+\tau^- -S}
\label{FS1}
\end{eqnarray}
 with
 \begin{eqnarray}
S=  \sum_{m,j}\frac{(-q^2)^j}{j!(2\mu+1)}\left(\frac{ \mu !}{m!}\right)^2 \left(\frac{\ell^+}{\beta^+}\right)^{2\mu}   \prod_{i=1}^{m} [k^2+(2i-1)q^2]. \nonumber\\
\label{FS2}
\end{eqnarray}

\end{document}